\documentclass[12pt, one column,comsoc]{IEEEtran}
\usepackage[T1]{fontenc}% optional T1 font encoding

\ifCLASSINFOpdf
\else
\fi
\usepackage{amsmath}
\usepackage[cmintegrals]{newtxmath}
\usepackage{algorithm}
\usepackage{algorithmic}
\usepackage{cite}
\usepackage{optidef}
\usepackage{xcolor}

\begin{document}
%
% paper title
% Titles are generally capitalized except for words such as a, an, and, as,
% at, but, by, for, in, nor, of, on, or, the, to and up, which are usually
% not capitalized unless they are the first or last word of the title.
% Linebreaks \\ can be used within to get better formatting as desired.
% Do not put math or special symbols in the title.
\title{Cross Layer Design for Maximizing Network Utility in Multiple Gateways Wireless Mesh Networks}
%
%
% author names and IEEE memberships
% note positions of commas and nonbreaking spaces ( ~ ) LaTeX will not break
% a structure at a ~ so this keeps an author's name from being broken across
% two lines.
% use \thanks{} to gain access to the first footnote area
% a separate \thanks must be used for each paragraph as LaTeX2e's \thanks
% was not built to handle multiple paragraphs
%

\author{Samaneh~Aghashahi,~\IEEEmembership{ Student Member,~IEEE,}
        Ghasem~Mirjalily,~\IEEEmembership{Senior Member,~IEEE,}
        \\ Aliakbar~Tadaion,~\IEEEmembership{Senior Member,~IEEE}
}

\maketitle

% As a general rule, do not put math, special symbols or citations
% in the abstract or keywords.
\begin{abstract}
We investigate the problem of network utility maximization in multiple gateways wireless mesh networks by considering Signal to Interference plus Noise Ratio (SINR) as the interference model. The aim is a cross layer design that considers joint rate control, traffic splitting, routing, scheduling, link rate allocation and power control to formulate the network utility maximization problem. As this problem is computationally complex, we propose the Joint dynamic Gateway selection, link Rate allocation and Power control (JGRP) algorithm based on the differential backlog as a sub-optimal solution. This algorithm first constructs the initial network topology, and then in each time slot, determines the  generation rate and destination gateway of each traffic flow, simultaneously.
The other main task of this algorithm is joint routing, scheduling, links rate allocation and node power allocation in each time slot. Moreover, for improving the fairness, we propose  some new parameters instead of the differential backlog in JGRP algorithm. Simulation results show that using the proposed parameters in JGRP algorithm improves fairness from throughput and delay point of views.
\end{abstract}

% Note that keywords are not normally used for peerreview papers.

% For peer review papers, you can put extra information on the cover
% page as needed:
% \ifCLASSOPTIONpeerreview
% \begin{center} \bfseries EDICS Category: 3-BBND \end{center}
% \fi
%
% For peerreview papers, this IEEEtran command inserts a page break and
% creates the second title. It will be ignored for other modes.
\IEEEpeerreviewmaketitle

\section{Introduction}

We study the  network utility maximization problem  by jointly considering rate control, traffic splitting among gateways, routing, scheduling, link rate allocation and power control in multiple gateways wireless mesh networks. Over the past two decades, the mesh structure has been considered as an appropriate solution to increase the coverage area and capacity of wireless networks\cite{WMN}. Important features of the wireless mesh networks include low cost deployment, distributed communication and robustness. However, the performance of these networks could be degraded, which is mainly due to poor design of network protocols  \cite{wirelessurv,survc}.

In recent years, various approaches have been provided to improve the performance of wireless mesh networks, among them is cross layer design which could be performed with various aims, such as improvement of throughput, delay and other network parameters. Another approach is using multiple gateways in these networks. In the following, we briefly review some related works according to these approaches.

First, some researches on cross layer design in wireless mesh networks. 
 In \cite{Nik}, the authors investigated joint routing, channel assignment, power control and rate adaptation to improve the throughput, load balancing and fault-tolerant in multi-radio multi-channel wireless mesh networks. As this problem is NP-hard, they proposed a heuristic algorithm with two levels. In the first level, a $K$-connectivity network topology is created using channel assignment and routing. In the second level, power control, rate adaption and scheduling are jointly considered  for maximizing the throughput while $K$-connectivity network topology is preserved. In order to maximize the capacity of multi-radio mult-channel wireless mesh networks, the authors in \cite{LRA}, considered the link rate allocation, routing and channel assignment. In \cite{Algo-PolyH}  the scheduling and routing design are performed jointly with the aim of minimizing the superframe length to support any random demand in multi-Tx/Rx wireless mesh networks. The authors of \cite{delay2} considered joint scheduling and routing in multi-Tx/Rx wireless mesh networks for minimizing the end-to-end delays and superframe length. In \cite{pso}, joint optimization of channel assignment, power control and routing  is investigated under the  Signal to Interference plus Noise Ratio (SINR)   model with the aim of increasing the network capacity. As this joint optimization problem is NP-hard, the Genetic 
 and particle swarm optimization algorithms are employed in \cite{pso} for optimizing channel assignment and power control, and then according to the optimal values obtained by these two algorithms, optimal routing is achieved by solving an LP problem . In \cite{multihop1} the authors designed a joint routing and power control mechanism for reducing the power consumption in large wireless mesh networks . The authors of \cite{multicast3} considered joint routing and channel assignment to do multiple multicast routing  and  showed that this design increases the network throughput. In \cite{multihop2}, joint routing and power control are considered to make trade-off between delay and energy consumption in wireless mesh networks. In \cite{cooperative}, the authors considered joint scheduling and channel assignment to increase the throughput and load balancing of  multi-radio multi-channel wireless mesh networks. The authors in \cite{CDC,CLC} have proposed joint rate control and scheduling for increasing the network utility. In \cite{Weighted-Backpressur}, for improving the quality of service parameters such as reliability and end-to-end delay, the authors proposed a joint scheduling and routing algorithm. In \cite{Directional}, the authors considered joint rate control, routing, channel assignment and scheduling to maximize the network utility of the multi-radio multi-channel wireless mesh networks with directional antennas. As the considered problem in \cite{Directional} is mixed integer nonlinear problem (MINLP), the authors used generalized Benders decomposition approach to solve it. In \cite{traffic}, the authors investigated joint power allocation and channel assignment for maximizing the aggregate throughput of cognitive wireless mesh networks. In \cite{cross2018}, resource allocation scheduling and routing are jointly determined  to maximize the  network utility of  wireless mesh networks in cloud computing. In \cite{jointtopology}, the authors considered joint topology control and partially overlapping channel assignment to improve the capacity of multi-radio multi-channel wireless mesh networks.

As mentioned before, a solution to improve the performance of wireless mesh networks is considering multiple gateways for these networks. In \cite{plasma}, a heuristic routing algorithm is proposed to increase the network throughput. This algorithm determines the transmission  rate and destination gateway of each flow. In \cite{MGMR} the authors considered multi-rate multicast routing in multiple gateways multi-radio multi-channel wireless mesh networks for maximizing the throughput.Then, the authors split this NP-hard problem into three phases: gateway selection, channel assignment and rate allocation.
 In \cite{costff}, considering multiple gateways, the authors proposed a multicast routing algorithm which constructs a multicast tree by maximizing the multicast-tree transmission ratio, and they showed that this algorithm improves the average delay and delivery ratio . In \cite{optimization} the authors considered the problem of multicast routing with multiple gateways and partially overlapped channels, and they showed that such techniques in this problem lead to reduce the links interference.

The authors of \cite{multigate} employed both cross layer design and multiple gateways approaches to improve the performance of wireless mesh networks. The authors considered joint rate control, traffic splitting, routing and scheduling under one-hop interference model to maximize the network utility of multiple gateways wireless mesh networks and they showed that using both cross layer design and multiple gateways approaches considerably improves the throughput and fairness.

In this paper, we consider joint rate control, traffic splitting, scheduling, routing, link rate allocation and power control under SINR as the interference model in a multiple gateways wireless mesh network. Actually, by considering the SINR  model, we investigate a more realistic scenario compared to  \cite{multigate}, which has considered the one hop interference model. In addition, besides rate control, traffic splitting, routing and scheduling that has been considered in  \cite{multigate}, we consider also link rate allocation and power control in our cross layer design, as these tools have important roles in SINR model.  Similar to \cite{multigate}, our aim is maximizing  the network utility which is a widely-used performance metric and could measure both the aggregated throughput and fairness in the network.  In this paper, we propose Joint dynamic Gateway selection, link Rate allocation and Power control  (JGRP) algorithm based on the differential backlog  as  a sub-optimal solution for solving the network utility maximization problem. This algorithm has three parts; in the first part, the network topology is formed by pruning the full mesh network to reduce the complexity of other parts. In the second part, the mechanisms of rate control and traffic splitting  are jointly obtained and in  the third part, joint scheduling, routing, rate allocation to links and power allocation to nodes are obtained  by employing a sub-optimal search  method which we present. Moreover, we propose some new parameters instead of the differential backlog to improve the fairness of our JGRP algorithm.

The rest of the paper is organized as follows: In Section \ref{sect2}, we describe the network model.  In Section \ref{sect3}, the network utility maximization problem is  formulated.  Section \ref{sect4} describes  the proposed  JGRP  algorithm as a sub-optimal solution to solve the network utility maximization problem. In Section \ref{sect5}, we attempt to improve the fairness by defining some new parameters. We provide some simulation results in Section \ref{sect6}, and finally Section \ref{sect7} concludes the paper.
% You must have at least 2 lines in the paragraph with the drop letter
% (should never be an issue)

% needed in second column of first page if using \IEEEpubid
%\IEEEpubidadjcol
\section{Network model}
\label{sect2}
We consider a wireless mesh network, where we have $N$ mesh nodes and $L$ links. We model the network with directed graph $G=(\Gamma,E)$, where $\Gamma$ is the set of mesh nodes and $E$ is the set of links. We assume that there are multiple gateways in this networks  and $GW \subset \Gamma$ represents the set of mesh gateways.
\subsection{Interference  Model}
In order to model the interference, we consider SINR model, where two directed links $(p,q)$,$(i,j)$ could be activated with  $r_{ij}$ and $r_{pq}$ rates, simultaneously if and only if these links satisfy the following conditions:
\begin{equation}
\begin{split}
  & \frac{{G_{ij} P_{ij} }}{{N_0  +  {G_{pj} P_{pq} } }} \ge \beta(r_{ij} )  \quad (i,j) ,(p,q) \in  E \\
  &  \frac{{G_{pq} P_{pq} }}{{N_0  +  {G_{iq} P_{ij} } }} \ge \beta(r_{pq} )  \quad  (i,j) ,(p,q)  \in  E
   \end{split}
   \label{eq1}
\end{equation}
where $P_{ij}$ is the transmission power from node $i$ to node $j$, $\beta(r_{ij})$ is  the SINR threshold for acceptable bit error probability, and $N_0$ is the background noise power. Moreover, $G_{ij}$ denotes the channel gain between nodes $i$ and $j$ and  equals $G_{ij}=(\frac{d_{ij}}{d_0})^{-\alpha}$, where $d_{ij}$ is the distance between nodes $i$ and $j$, $d_0$ is the reference distance and $\alpha$ denotes the path  loss exponent.
\subsection{Scheduling, Link Rate Allocation and Power Control}
Considering the interference model, scheduling specifies which of the links could be activated, simultaneously. We represent the set of  feasible schedules with $\Phi$ and the vector $\mathbf{s}_{m}$ as the $m^{\rm th}$ feasible schedule in which the links with common nodes could not be activated. We denote the element of $\mathbf{s}_{m}$ corresponding to link $(i,j)$ with $Y_{ij}^m$ which is equal to one if the link $(i,j)$ is activated in this schedule and zero, otherwise. In addition, $\pi_m$ denotes the fraction of time slot when  $\mathbf{s}_{m}$ is activated.  Now, we extend equation \eqref{eq1} for a feasible schedule as follows:
 \begin{equation}
   \frac{{G_{ij} P_{ij}^{m} }}{{N_0  + \sum\limits_{\substack{(p,q) \in \mathbf{s} _m \\ (p,q)\neq (i,j)}}^{} {G_{pj} P_{pq}^{m}} }} \ge \beta(r_{ij}^{m} )  \hspace{.1cm}   \forall  (i,j) \in \mathbf{s} _m , \\ 
   \label{eq2}
\end{equation}
where the links may have different transmission rates and powers in nonidentical feasible schedules. Assuming that $Y_{ij}^m=1$, $r_{ij}^m$ is the allocated rate to link $(i,j)$ and $P_{ij}^m$ is the transmit power from node $i$ to node $j$ where the transmission rate is equal to $r_{ij}^m$.

\subsection{Traffic and Queueing Model}
\textbf{Traffic Model}: We assume that the number of traffic flows in our network is $K$, where $F=\{ 1,2,...,K\}$ represents the set of all traffic flows in the network. We show the acceptable amount of traffic for flow  $f$  at time slot $t$ by $r_{s(f)}^{(f)}(t)$, where $s(f)$ is  the source node of flow $f$ and considering $ \sum\nolimits_{f:s\left( f \right) = i} {r_i^{(f)}(t) }  \le R_i^{\max }    \forall  i \in \Gamma  $ as a constraint that shows the limitation of each node $i$ in generating traffic. We assume that each of the gateways could be chosen as the destination of the packets of each flow $f$, where $y_d^{(f)}$ shows the fraction of the traffic of  flow $f$ forwarded to gateway $d$.

\textbf{Queening Model}: We assume that there are multiple queues in each node, where each queue is corresponding to one of the gateways. The packets corresponding to a gateway lie in the same queue, even if they belong to different flows. We represent the length of the queue in node $i$ corresponding to gateway $d$ at the beginning of time slot $t$ by $Q_i^{(d)}(t)$, where the  queue of gateway $d$ corresponding to this gateway is assumed empty, i.e., $Q_d^{(d)}(t)=0$. In addition, we denote the number of  packets belong to destination $d$, which is  transmitted over link $(i,j)$  at time slot $t$ by $\mu_{ij}^{(d)}(t)$ and the long term average of this parameter  by $\mu_{ij}^{(d)}$. Moreover,  we represent all traffic on  link $(i,j)$ by $\mu_{ij}$. By these definitions, it is clear that the queue evolution is as follows \cite{rbook}:
\begin{equation}
\begin{split}
&Q_i^{(d)} \left( {t + 1} \right) \le \max \left[ {Q_i^{(d)} \left( t \right) - \sum\limits_{b \in \Gamma } {\mu _{ib}^{\left( d \right)} \left( t \right)}  , 0} \right]\\&+ \sum\limits_{f:s\left( f \right) = i} {y_d^{(f) }r_i^{(f )}\left( t \right)}  + \sum\limits_{a \in \Gamma } {\mu _{ai}^{\left( d \right)} \left( t \right)},
\label{eq3} 
\end{split}
\end{equation}
where $ \sum\limits_{b \in \Gamma } {\mu _{ib}^{\left( d \right)} \left( t \right)}$ is the output traffic from node $i$, $\sum\limits_{a \in \Gamma } {\mu _{ai}^{\left( d \right)} \left( t \right)}$ is the input traffic to node $i$ from upstream nodes and $\sum\limits_{f:s\left( f \right) = i} {y_d^{(f)} r_i^{(f )}\left( t \right)}$ is the generated traffic in node $i$.

\section{Problem Formulation}
\label{sect3}
Now, we formulate  the network utility maximization problem under the constraints corresponding to scheduling, rate control, link rate allocation and power control. We define $\mathbf{r}=[r_{s(1)}^{(1)},r_{s(2)}^{(2)},\dots,r_{s(K)}^{(K)}]^{T}$ as the long-term average traffic vector. The aim is to maximize of the sum of  long-term average  traffics of all  network flows. Moreover,  as we would like to have fairness among the flows,  $\rm{log(.)}$ function  is considered as the utility function of the problem, which is also considered in\cite{CDC,CLC,multigate},
\begin{maxi!}|s|[2]    
             % mini! = minimize 
    {\mathbf{r}}{\sum\limits_{f = 1}^K {\log \left( {r_{s(f)}^{(f )}} \right)} \label{cer4}}   % objective function and label
            % label for optimizatio problem
      {\label{eq4}}            
    {}                                % optimization result
  \addConstraint{\!\!\!\!\!\!\sum\limits_{f:s\left( f \right) = i}{y_d^{\left( f \right)} r_{s(f)}^{\left( f \right)} }  + \sum\limits_{a \in \Gamma} {\mu _{ai}^{\left( d \right)} } }{=\sum\limits_{b \in \Gamma} {\mu _{ib}^{\left( d \right)}}; \forall i \in \Gamma , d \in GW , i \neq d  \label{eq4-1}  } 
  % constraint 1
 %  \addConstraint
 \addConstraint{\sum\limits_{d \in GW}^{} {y_d^{\left( f \right)} }}{= 1 \label{eq4-2}}  % constraint 2
 \addConstraint{\mu _{ij}}={ \sum\limits_{d \in GW}^{} {\mu _{ij}^{\left( d \right)} }; \forall (i,j) \in E \label{eq4-3}} 
    \addConstraint{\sum\limits_{m = 1}^{\left| \Phi \right|} {\pi _m }}{= 1 \label{eq4-4}}    
    \addConstraint{\mu _{ij}}{= \sum\limits_{\left( {i,j} \right)  \in  \mathbf{s} _m } {\frac{r_{ij}^{m}  \times \pi _m  \times l_t }{l_p}}  ; \forall (i,j) \in E \label{eq4-5}}
         \addConstraint{\frac{{G_{ij} P_{ij}^{m} }}{{N_0  + \sum\limits_{\substack{(p,q) \in \mathbf{s} _m \\ (p,q)\neq (i,j)}}^{} {G_{pj} P_{pq}^{m} } }} }{\ge \beta(r_{ij}^{m} ) \hspace{.1cm}  ; \forall  (i,j) \in \mathbf{ s} _m  \label{eq4-6}}   
         \addConstraint{Y_{ij}^m}{\in \{0,1\}\label{eq4-7}}  
             \addConstraint{Y_{ij}^m+Y_{ji}^m\leq 1}{,Y_{ip}^m+Y_{pq}^m\leq 1,Y_{ip}^m+Y_{iq}^m\leq 1,Y_{ij}^m+Y_{pj}^m\leq 1\label{eq4-8}}
               \addConstraint{0<P_{ij}^m}{\leq P_{max} \label{eq4-9}}
\end{maxi!}
The constraint \eqref{eq4-1} states that the total input traffic rate to the queue of  gateway $d$ in node $i$ must be equal to the output traffic rate from this queue. 
 The constraint \eqref{eq4-2} ensures that the total traffic of flow $f$ is forwarded to the gateways.
 The constraint \eqref{eq4-3} indicates that the sum rate transmitted over link $(i,j)$ by different destinations must be equal to all traffic rate on the link $(i,j)$. The constraint \eqref{eq4-4} means that sum of the fractions of the time slot corresponding to various feasible schedules must be equal to one. The constraint \eqref{eq4-5} ensures that the allocated rate and the activation time of the link $(i,j)$ are enough for carrying the desired amount of traffic, where $l_t$ and $l_p$ represent the duration of each time slot and packet length, respectively. The constraint \eqref{eq4-8} shows that each of the nodes in each scheduling could not be connected to more than one of the other nodes. The constraint \eqref{eq4-9} limits the transmission power of the nodes.
\section{Sub-Optimal Solution}
\label{sect4}
We must note that problem \eqref{eq4} is NP-hard (see e.g. \cite{hard}), then in this section, we propose a sub-optimal solution to solve the network utility maximization problem \eqref{eq4}. Accordingly, we offer JGRP  algorithm which has three parts:
\begin{itemize}
\item 
In the first parts, we prune the full mesh network and form an initial network topology to reduce the computational complexity of the third part of  the algorithm.
\item
In the second part, we  jointly obtain the mechanism of the traffic splitting and rate control.
\item
In  the third part, we determine the routing, scheduling, rate allocation to the links and power allocation to the nodes, Simultaneously.
\end{itemize} 
We must note that the first part of the algorithm  runs at the beginning (before network operation) ; but other parts run repeatedly  at the beginning of each time slot.
\subsection{Construction of  Initial Network Topology }
Assume that the number of links that could be connected to each node is an integer number between   $\left[\frac{N}{3}\right]$ and $2\left[\frac{N}{3}\right] $, where $N$ is the number of the nodes. Considering the channels between the nodes, we propose Algorithm \ref{algo1} to form the initial network topology. 
\begin{algorithm}[h]
\caption{Constructing the initial network topology.  }
\label{algo1}
\begin{algorithmic}[1]
\REQUIRE
The set of the mesh nodes $\Gamma$, The channel gains between  nodes $i,j \in \Gamma$  $G_{ij}$.
\STATE
For each  node $i$,  arrange the other nodes decreasingly, according to the channel gains between each of the nodes and node $i$.
\STATE
Considering the priorities of  all the nodes based on ordering in step $1$, establish a link between the nodes which are in the first $\left[\frac{N}{3}\right]$ priority of each other.
\FOR{$k=\left[\frac{N}{3}\right]$ to $k=2\left[\frac{N}{3}\right]-1$}
\FOR{$j=1$ to $ j=\left[\frac{N}{3}\right]$}
\STATE
For each node  $i$ which is not connected to  its  $j^{\rm{th}}$ priority node, if the number of the links of the $j^{\rm{th}}$ priority node of node $i$ is equal to $k$, establish a link between this node and its $j^{\rm{th}}$ priority node.
\ENDFOR
\ENDFOR
\end{algorithmic}
\end{algorithm}

\subsection{Rate Control and Traffic Splitting}
 Similar to  \cite{multigate}, at the beginning of each time slot $t$, the rate controller at the source node $s(f)$ of each flow $f \in F$ selects the gateway with the shortest queue length and enters the traffic to the queue as follows:
\begin{equation}
  r_{s\left( f \right)}^{*\left( f \right)}(t)  = \frac{V}{{Q_{s\left( f \right)}^{\left( {d^* } \right)} \left( t \right)}},
  \label{eq5}
\end{equation}
where $V$ is  a constant parameter  which controls the trade-off between the network utility and the queue length, $d^*$ is the gateway which has the shortest queue in $s(f)$ and $Q_{s\left( f \right)}^{\left( {d^* } \right)} \left( t \right)$ is the queue length of $d^*$ in $s(f)$.
\subsection{Routing, Scheduling, Link Rate Allocation  and Power Control}
For each link $(i,j)$, the differential backlog is defined as,
 \begin{equation}
W_{ij}=\mathop {\max }\limits_{d \in  GW} \left[ {Q_i^{\left( d \right)} \left( t \right) - Q_j^{\left( d \right)} \left( t \right)} \right ]
 \label{eq6}
 \end{equation}
This parameter specifies the gateways whose traffic is carried over link $(i,j)$. In addition, the differential backlog is related to  the amount of congestion at nodes $i$ and $j$.
 
We must note the existed congestion in  the nodes and maximization of the network links throughput during the specification of routing, scheduling and the rate allocation  to the links and  the power allocation  to  the nodes. In this end, we involve $W_{ij}$, $\pi_m$ and $r_{ij}^m$ in the objective function of the problem to consider the amount of congestion of the nodes and  the   throughput of the links. So, we formulate the problem as follows:
\begin{equation}
 \begin{split}
&\mathop {\max }\limits_{r_{ij}^{m},P_{ij}^{m},\pi _m ,Y_{ij}^m}  \sum\limits_{m:\mathbf{ s} _m  \in \Phi } {\sum\limits_{(i,j) \in E} {(\frac{r_{ij}^{m}  \times \pi _m  \times l_t}{l_p } + W_{ij} )Y_{ij}^m } } \\
&\mathrm{s.t} \\
 &  \sum\limits_{m = 1}^{\left| \Phi \right|} {\pi _m }  = 1 \\ 
        & \frac{{G_{ij} P_{ij}^{m} }}{{N_0  + \sum\limits_{l(m,n) \in \mathbf{s} _m }^{} {G_{pj} P_{pq}^{m} } }} \ge \beta(r_{ij}^{m} ) \hspace{.1cm}   \forall  (i,j) \in \mathbf{ s} _m \\ 
    &  Y_{ij}^{m} \in \{0,1\} \\
&{Y_{ij}^m+Y_{ji}^m\leq 1}{,Y_{ip}^m+Y_{pq}^m\leq 1,Y_{ip}^m+Y_{iq}^m\leq 1,Y_{ij}^m+Y_{pj}^m\leq 1} \\
   & 0 < P_{ij}^{m}     \le P_{\max }    
    \label{eq7}
\end{split}
 \end{equation}
 In order to solve the problem \eqref{eq7}, we first assume that the variable $\pi_m$  be constant and reformulate the problem as follows:
  \begin{equation}
 \label{eq8}
 \begin{split}
&\mathop {\max }\limits_{r_{ij}^{m},P_{ij}^{m},\pi _m ,Y_{ij}^m}  \sum\limits_{m: \mathbf{ s} _m  \in \Phi } {\sum\limits_{(i,j) \in E} {(\frac{r_{ij}^{m}  \times \pi _m  \times l_t}{{l_p} } + W_{ij} )Y_{ij}^m } }  \\
&\mathrm{s.t} \\
        & \frac{{G_{ij} P_{ij}^{m} }}{{N_0  + \sum\limits_{l(m,n) \in \mathbf{ s} _m }^{} {G_{pj} P_{pq}^{m} } }} \ge \beta(r_{ij}^{m} ) \hspace{.1cm}   \forall  (i,j) \in \mathbf{ s} _m \\ 
    &  Y_{ij}^{m} \in \{0,1\} \\
&{Y_{ij}^m+Y_{ji}^m\leq 1}{,Y_{ip}^m+Y_{pq}^m\leq 1,Y_{ip}^m+Y_{iq}^m\leq 1,Y_{ij}^m+Y_{pj}^m\leq 1} \\
   & 0 < P_{ij}^{m}     \le P_{\max }    
\end{split}
 \end{equation}

  In problem \eqref{eq8} the variables  $r_{ij}^m$ and $\beta(r_{ij}^m)$ should be selected from the discrete set of allowable rates  and the variable $Y_{ij}^m$ is binary. Accordingly, we need a full search to obtain the optimal solution, which is not practical. Hence, we propose a sub-optimal search (Algorithm \ref{algo3}) to solve the problem. As a prerequisite, we rewrite the SINR constraints corresponding to a typical set of links which have not any common nodes denoted by $ S\subset E$. In
other words, for all values of $(i_k,j_k) \in S$ and $(i_n,j_m) \in S$, we have $ i_m\neq i_n,i_m\neq j_n ,j_m\neq i_n,j_m\neq j_n $.
 The SINR constraint for link $(i_k,j_k)$ would be,
 \begin{equation}
 \frac{{G_{i_kj_k} P_{i_kj_k}^{m} }}{{N_0  + \sum\limits_{ \substack{(i_n,j_n) \in \mathbf{S} \\ n\neq k}}^{} {G_{i_kj_n} P_{i_nj_n}^{m} } }} \ge \beta(r_{i_kj_k}^{m})
 \label{eq10}
\end{equation}
By replacing the inequality to equality in the above constraint, we have:
  \begin{equation}
\frac{1}{\beta(r_{i_kj_k}^{m})}( G_{i_kj_k}P_{i_kj_k}^{m})- \sum\limits_{ \substack{(i_n,j_n) \in \mathbf{S} \\ n\neq k}}^{} {G_{i_kj_n} P_{i_nj_n}^{m}}=N_0 
\label{eq11}
  \end{equation}
  Now, by writing \eqref{eq11} for all  links, we have:
    \begin{equation}
     \label{eq12}
  \begin{bmatrix}
P_{i_1,j_1}^m\\
P_{i_2,j_2}^m\\
\vdots
\\
P_{i_n,j_n}^m\\
\vdots\\
P_{i_{M},j_{M}}^m
\end{bmatrix}=\mathbf{A}^{-1}\mathbf{1}N_0
\end{equation}
  Where matrix   $\mathbf{A}$ is:
   \begin{equation}
 \label{eq13}
 \mathbf{A}=
\begin{bmatrix}
\frac{1}{\beta(r_{i_1j_1}^{m})} G_{i_1j_1}  & \hdots &  -G_{i_nj_1} & \hdots & -G_{i_Mj_1} 
\\
-G_{i_1j_2}   & \hdots &  -G_{i_nj_2} & \hdots & -G_{i_Mj_2}
\\
\vdots
\\
-G_{i_1j_n} & \hdots & \frac{1}{\beta(r_{i_nj_n}^{m})} G_{i_nj_n} & \hdots &  -G_{i_Mj_n}
\\
\vdots
\\
-G_{i_1j_M} & \hdots &  -G_{i_nj_M} & \hdots &  \frac{1}{\beta(r_{i_Mj_M}^{m})} G_{i_Mj_M} 
\end{bmatrix}
 \end{equation} 
\begin{algorithm}
% \algsetup{linenosize=\small}
  %\scriptsize
\caption{Obtaining joint scheduling, routing, power control and link rate allocation.}
\label{algo3}
\begin{algorithmic}[1]
\REQUIRE
The graph of the network the channels between the nodes  $i$ and $j$ $\forall i,j$, The set of rates in the Standard $802.11\mathrm{a}$;
$R=\{R_1, R_2, \dots, R_8 |R_1>R_2\dots >R_8\}$, the set of  SINRs corresponding to the rates in the set $R$; $\beta=\{\beta(R_1),\dots,\beta(R_8)\}$,  the length of queue corresponding to gateway $d$ in node $i$; $Q_{i}^{d}(t)$. 

\ENSURE
The activated schedules, the rates of the links  and the power of the nodes, in each time slot.
    
\STATE 
For each link
$ (i,j) \in E$,
compute the differential backlog as:
 \begin{equation*}
W_{ij}=\mathop {\max }\limits_{d \in  GW} \left[ {Q_i^{\left( d \right)} \left( t \right) - Q_j^{\left( d \right)} \left( t \right)} \right ].
 \end{equation*}

\STATE
Sort 
$W_{ij}$s
decreasingly and lie them in the first row of 
$\mathbf{W}^{sort} \in \mathbb{R}^{3 \times L}$,
and 
set the second and third row of the column of $\mathbf{W}^{sort}$ corresponding to $W_{ij}$,
as 
$[i,j]^T$.
   \vspace{.1cm} 
\STATE
Select the first $M$ column of  $\mathbf{W}^{sort}$ and sort the second and third row of them as ordered pair and lie them in the sets  $\mathbf{sch}_1\dots \mathbf{sch}_M$, respectively. Moreover, allocate the rate  $R_1$ to these links and lie them in  the families   $\mathbf{R}_{s_1}\dots \mathbf{R}_{s_M}$, respectively.
   \vspace{.1cm} 
\FOR{   $n=1$
to
  $n=|R|$}
     \vspace{.1cm} 
  \FOR{  $q=L-M$
  to
  $q=L$}
   \vspace{.1cm} 
  \IF{  
  $[\mathbf{W}^{sort} ]_{1q}>=0$
 and
  $[W_1]_q=0$}
  \FOR{$m=1$ to $m=M$}
     \vspace{.1cm} 
 \IF{
 If the link corresponding to $ [\mathbf{W}^{sort} ]_{1q}$ do not have common node with any links of  the set  $\mathbf{sch}_m$}
    \vspace{.1cm} 
 \STATE set 
  $\mathbf{temp}=(i,j)\cup \mathbf{sch}_m$
   \vspace{.1cm} 
  \STATE
  Set the rate of the link corresponding to $ [\mathbf{W}^{sort} ]_{1q}$, as $R_n$,
 set SINR of this link as $\beta(R_n)$ and set 
  $\mathbf{temp_r}=\{R_n\} \cup  \mathbf{R}_{s_m} $.
  
 \vspace{.2cm} 
  \STATE
  Considering the saved rates in $\mathbf{temp_r}$, formulate   problem \eqref{eq12} for
  $\mathbf{temp}$ and compute the powers; If  the power of all nodes in $\mathbf{temp}$ were positive and  smaller than $P_{max}$, set
  $r_{ij}^{m}=R_n$,
$\mathbf{R}_{s_m}=\mathbf{temp_r}$,
$\mathbf{sch}_m=\mathbf{temp}$
and
  $[W_1]_q=1$.
  \vspace{.1cm}
  \ENDIF
  
  \ENDFOR
  \ENDIF
  \ENDFOR
  \ENDFOR

\FOR{$m=1$
to
$m=M$}
\STATE 
In order to form  $\mathbf{s}_m$, if  any link $(i,j)$ is in $\mathbf{sch}_m$, set    $Y_{ij}^m=1$, else set it as zero.
\ENDFOR

 \STATE
Set
 $\Phi=[\mathbf{s}_1, \mathbf{s}_2 ,\dots \mathbf{s}_M]^T$.
 \label{algo20-3}
 \STATE
Find  $\pi_1,\dots,\pi_M$ by solving the following problem:
   \begin{equation*}
  \begin{split}
&\mathop {\max }\limits_{\pi_1,\dots,\pi_M}  \sum\limits_{m:\mathbf{s} _m  \in \Phi } {\sum\limits_{(i,j) \in E} {\frac{r_{ij}^{m}  \times \pi _m  \times l_t}{{l_p} } Y_{ij}^m } } \\
& \mathrm{s.t} \sum\limits_{m = 1}^{\left| \Phi \right|} {\pi _m }  = 1 
\end{split}
  \end{equation*} 
\end{algorithmic}
\end{algorithm}

Then, in Algorithm \ref{algo3} after solving problem \eqref{eq8}, in finally step of this algorithm, we  obtain $\pi_1,\dots,\pi_M$ by solving the following problem:s
   \begin{equation}
  \label{eq9}
  \begin{split}
&\mathop {\max }\limits_{\pi _m }  \sum\limits_{m:\mathbf{s} _m  \in \Phi } {\sum\limits_{(i,j) \in E} {\frac{r_{ij}^{m}  \times \pi _m  \times l_t}{l_p } Y_{ij}^m } } \\
& \mathrm{s.t} \sum\limits_{m = 1}^{\left| \Phi \right|} {\pi _m }  = 1 
\end{split}
  \end{equation}
\section{Improvement of the Fairness}
\label{sect5}
In Algorithm \ref{algo3} for link rate allocation and finding the active feasible scheduling in each time slot, we only considered the congestion of the nodes and the total rate of the activated links in each scheduling. Moreover, we allocated  more rate to the links with larger $W_{ij}$. This approach does not lead  to an acceptable fairness among the flows from the average delay and throughput  point of views. In this section, in order to improve the fairness among  the traffic flows, we define some new
parameters to be used in Algorithm \ref{algo3} instead of $W_{ij}$. These parameters are explained in the following,
\begin{enumerate}
\item  
Let $R_{ij}^{(T)}(t)$ be sum of the allocated  rates to link $(i,j)$ until the beginning of time slot $t$ and define $W_{ij}^{(r)}$ as:
\begin{equation}
W_{ij}^{(r)}=\frac{W_{ij}}{R_{ij}^{(T)}(t)}.
\label{eq15}
\end{equation}
By using  $W_{ij}^{(r)}$, the proposed algorithm allocates higher rates to the links to which we allocated  less rate before the beginning of time slot $t$.
\item 
Let $D_i^{(p)}(t)$ be the delay of the first packet lied in the queue of the gateway corresponding to $W_{ij}$  in time slot $t$. In each time slot $t$, we define $W_{ij}^{(D)}$ as,
\begin{equation}
W_{ij}^{(D)}=D_{i}^{(p)}(t)W_{ij}
\label{eq16}
\end{equation}
By using $W_{ij}^{(D)}$, we increase the probability of sending the packets which have experienced more delay.
\item
 In order to have trade-off between the packets delay   and the allocated  rate to  links, we define $W_{ij}^{(r,D)}$  as,
 \begin{equation}
W_{ij}^{(r,D)}=\frac{D_{i}^{(p)}(t)}{R_{ij}^{(T)}(t)}W_{ij}.
\label{eq5-17}
\end{equation}
\item
Assume  $d \in GW$ be the gateway corresponding to $W_{ij}$ and $R_{ij}^{(T,d)}(t)$ be the  sum  of  allocated rates to link $(i,j)$ for transmitting the traffic of  gateway $d$ until the beginning of time slot $t$. We define $W_{ij}^{(r_d)}$ as,
\begin{equation}
W_{ij}^{(r_d)}=\frac{W_{ij}}{R_{ij}^{(T,d)}(t)}.
\label{eq18}
\end{equation}
\item
In order to jointly consider  all of the previously defined
parameters related to delay,  allocated rate to the links and  allocated  rate to transmit the traffic of the gateways, we define $W_{ij}^{(r_d,D)}$ as,
\begin{equation}
W_{ij}^{(r_d,D)}=\frac{D_{i}^{(p)}(t)}{R_{ij}^{(T,d)}(t)}W_{ij}
\label{eq19}
\end{equation}
\end{enumerate}
\section{Simulation Results}
\label{sect6}
In order to evaluate and compare the proposed algorithms some simulation results are provided.  the algorithms are implemented using MATLAB. In all  simulations, we assume eight flows and two gateways. We set the simulation parameters as shown in Tables \ref{tab1} and \ref{tab2}.
\begin{table}[h]
 \begin{center}
 \caption{Simulation parameters }
  \label{tab1}
 \begin{tabular}{ll}
 \hline 
$N_\mathrm{0}$
power of background noise \cite{link} &  $-90$ {dBm} \\ 
  \hline 
 $\alpha$ path loss exponent \cite{powercontrol}
 & $3$\\
  \hline
   $d_\mathrm{0}$ reference distance
 & $10$ {m}\\
  \hline
duration of time slot \cite{crossarchitecture}& $625$ $\mu s$ \\ 
  \hline 
 packet length \cite{crossarchitecture}& $1470$ {bytes} \\ 
 \hline 
 $V$\cite{multigate} &$30 $
 \\
 \hline
$R_i^{\max} $ \cite{multigate}&$10$
 \\
 \hline
 $P_{max} $&$20$ {dBm}
 \\
 \hline
 \end{tabular} 
 \end{center}
 \end{table}
 \begin{table}[h]
  \begin{center}
  \caption{sinr thresholds required for supported rates in IEEE802.11a std\cite{802.11a}.}
  \label{tab2}
 \begin{tabular}{cc}
 \hline 
 Rate(Mbps) & SINR Threshold(dB)\\ 
 \hline 
$54 $& $24.56$ \\ 
 \hline 
$ 48$ & $24.05$ \\ 
 \hline 
 $36$ & $18.8$ \\ 
 \hline 
 $24$ & $17.04$ \\ 
 \hline 
 $18$ &$10.79$ \\ 
 \hline 
 $12$ & $9.03$ \\ 
 \hline 
 $9$ & $7.78$ \\ 
 \hline 
$ 6$& $6.02$\\ 
 \hline 
 \end{tabular} 
\end{center}
\end{table}

 \subsection{Investigation of the performance of JGRP  algorithm}
\begin{figure}[b]
 \centering
\includegraphics*[width=.5\columnwidth]{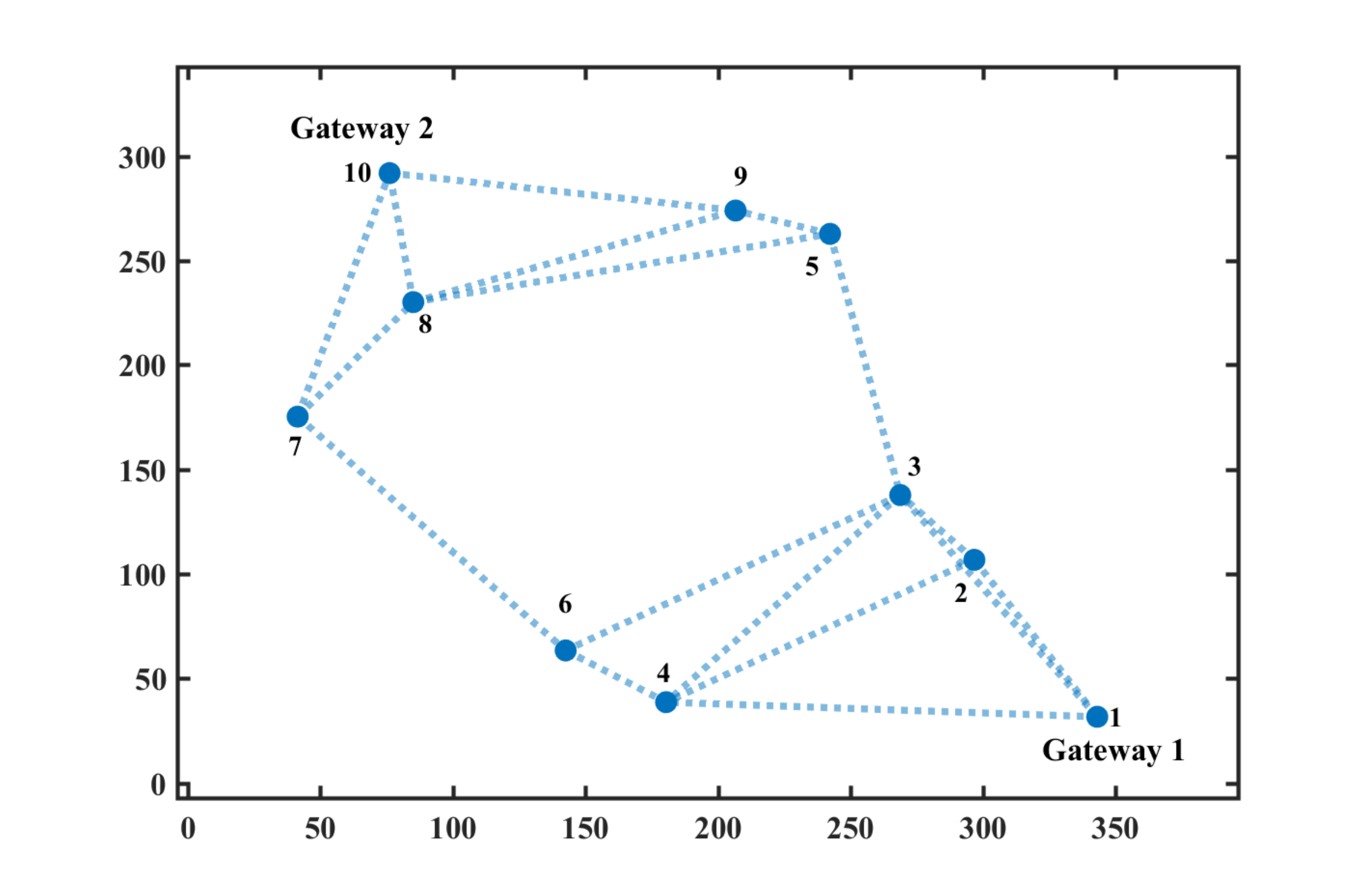}
\caption{The initial network topology constructed by Algorithm \ref{algo1}  in which nodes $1$ and $10$ are mesh gateways, and nodes $2$ to $9$ are the source nodes of the flows}
\label{fig1}
\end{figure}

\begin{figure}[h]
 \centering
\includegraphics*[width=.5\columnwidth]{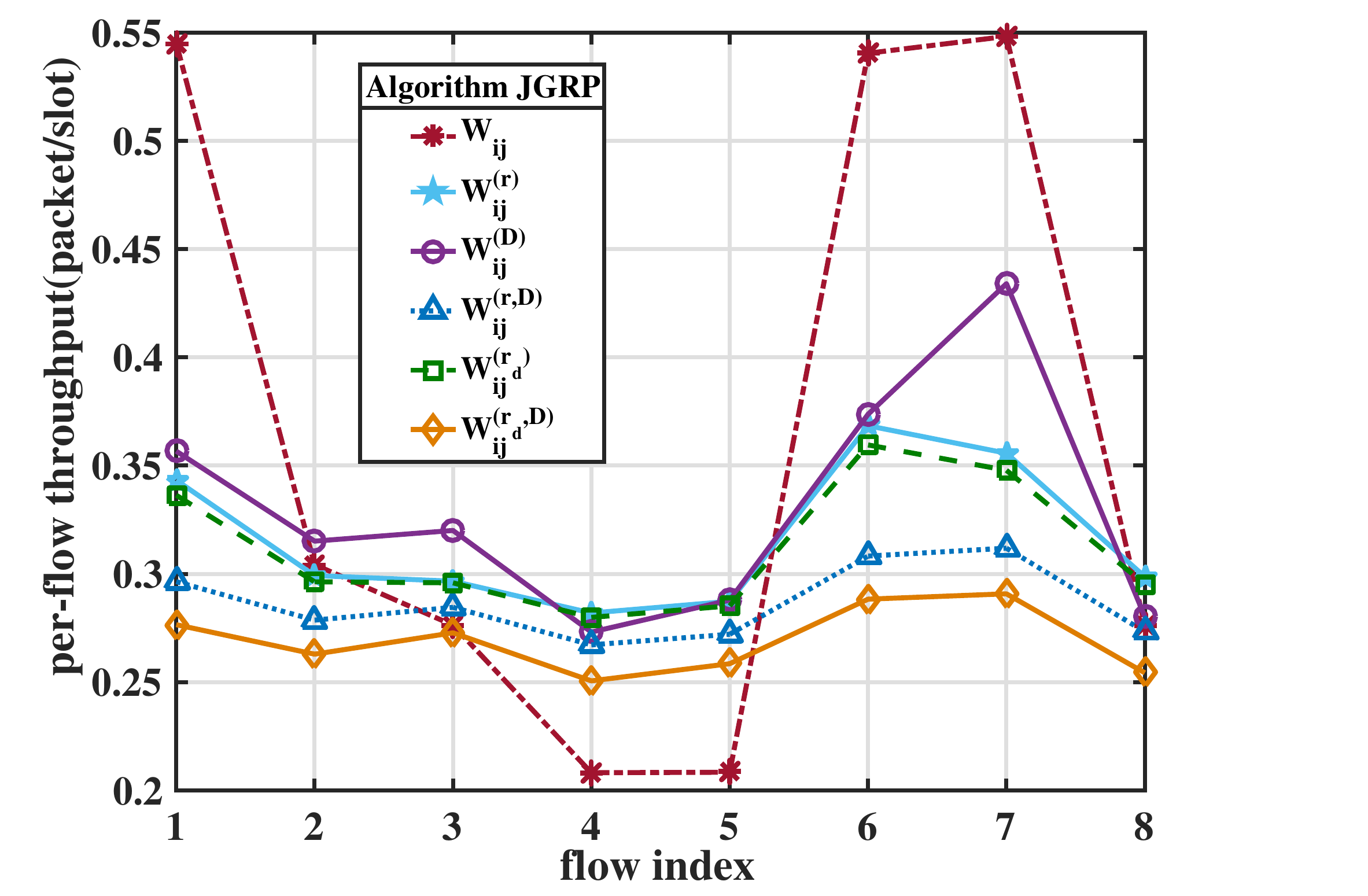}
\caption{Throughput of each flow in  JGRP  algorithm based on  parameters  $W_{ij}$, $W_{ij}^{(r)}$, $W_{ij}^{(D)}$,  $W_{ij}^{(r,D)}$, $W_{ij}^{(r_d)}$ and $W_{ij}^{(r_d,D)}$. }
\label{fig2}
\end{figure}

\begin{figure}[h]
 \centering
\includegraphics*[width=.5\columnwidth]{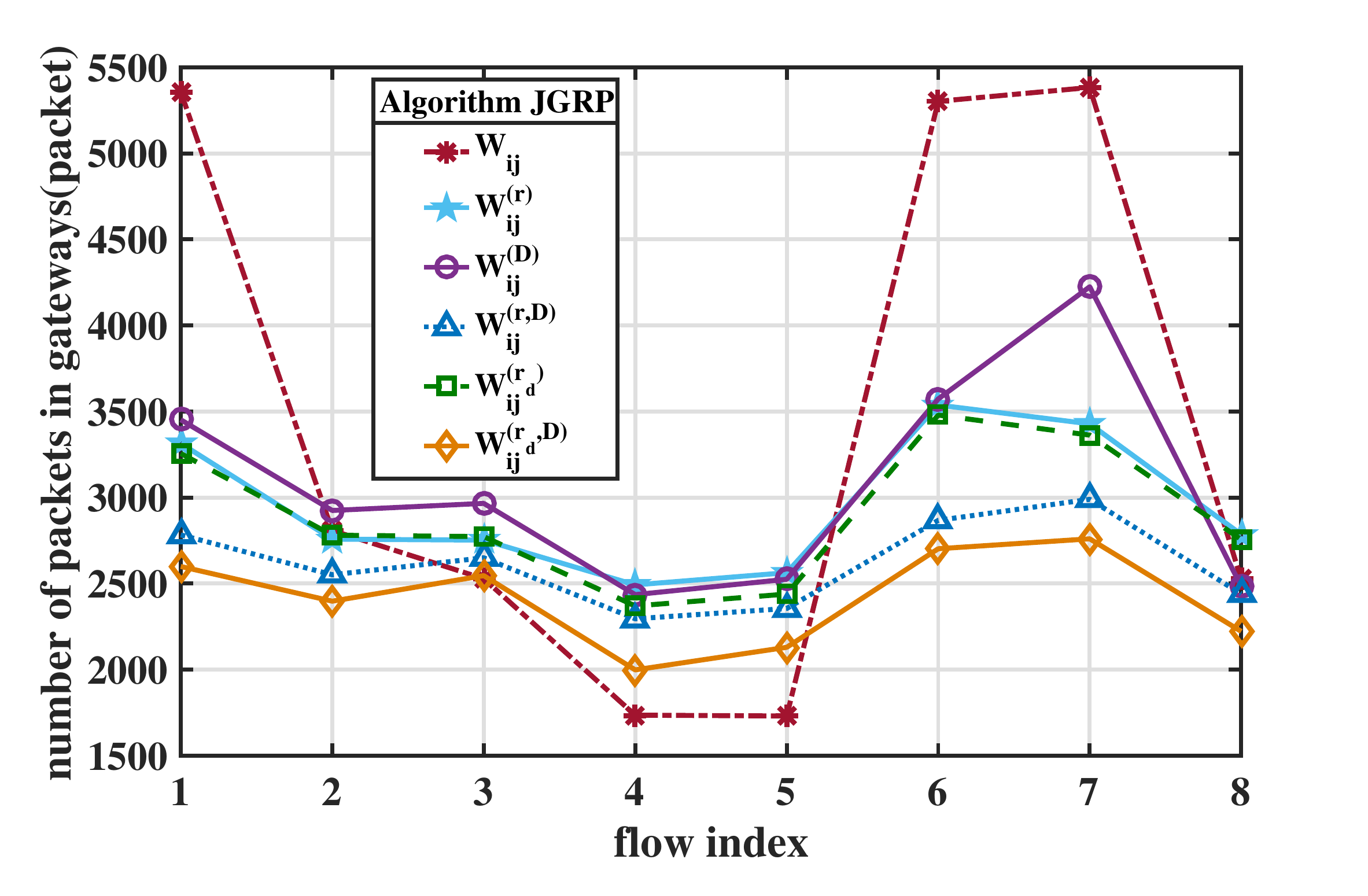}
\caption{Number of packets received by  the gateways of each flows in  JGRP  algorithm based on  parameters  $W_{ij}$, $W_{ij}^{(r)}$, $W_{ij}^{(D)}$,  $W_{ij}^{(r,D)}$, $W_{ij}^{(r_d)}$ and $W_{ij}^{(r_d,D)}$.}
\label{fig3}
\end{figure}

\begin{figure}[h]
 \centering
\includegraphics*[width=.5\columnwidth]{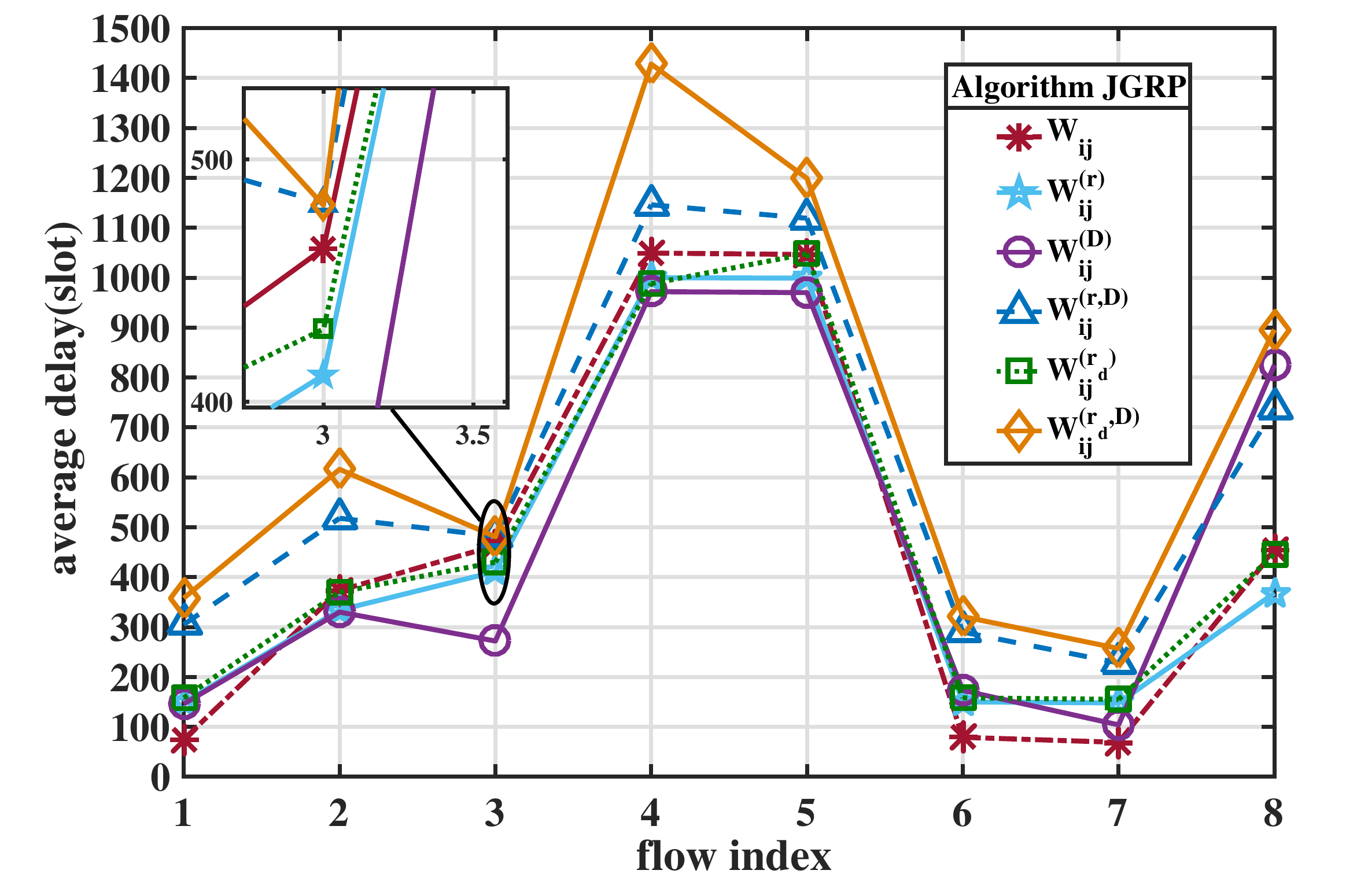}
\caption{Average delay of one packet for each flow in  JGRP  algorithm based on  parameters  $W_{ij}$, $W_{ij}^{(r)}$, $W_{ij}^{(D)}$,  $W_{ij}^{(r,D)}$, $W_{ij}^{(r_d)}$ and $W_{ij}^{(r_d,D)}$.}
\label{fig4}
\end{figure}

In the simulated network, $10$ nodes are distributed in $350m\times350m$ square area uniformly. By running Algorithm \ref{algo1}, the topology is constructed as shown in Fig. \ref{fig1}. We select two nodes $1$ and $10$ (which have the maximum distance from each other) as the mesh gateways, and we assume that  nodes $2$ to node $9$  are the sources of traffic flows  numbered $1$ to $8$, respectively. We run  JGRP   algorithm based on the six parameters   
$W_{ij}$
  ,
    $W_{ij}^{(r)}$
   ,
       $W_{ij}^{(D)}$
      ,
           $W_{ij}^{(r,D)}$ 
        ,
         $W_{ij} ^{(r_d)}$  
           and
        $W_{ij}^{(r_d,D)}$  on the constructed topology (Fig. \ref{fig1}) over $10^4$ time slots and compare their simulation results in Figs. \ref{fig2}, \ref{fig3} and \ref{fig4} and Tables \ref{tab3}, \ref{tab4} and \ref{tab5}.

In Fig. \ref{fig2}, we observe that by using the  JGRP  algorithm  based on $W_{ij}$, the  throughput  of different flows  are different and therefore there is no fairness among  flows. Moreover, we observe that this algorithm  based on $W_{ij}^{(r)}$,
       $W_{ij}^{(D)}$
       and
            $W_{ij}^{(r_{d})}$ improves the throughput  of  flows, that have low throughput  using the algorithm based on  $W_{ij}$. Furthermore, we observe that  using the  JGRP  algorithm based on $W_{ij}^{(r,D)}$    and  $W_{ij}^{(r_d,D)}$, the throughput of  
            various flows  are close to each other. This  is because of the reduction in throughput of the flows which achieve more throughput using  parameters $W_{ij}^{(r)}$, $W_{ij}^{(D)}$ and $W_{ij}^{(r_d)}$.

In Fig. \ref{fig3}, we observe that by using  JGRP  algorithm based on $W_{ij}$, the number of packets corresponding to various flows received by the gateways are very different. Moreover, it can be observed that using  JGRP  algorithm based on  $W_{ij}^{(r)}$
    ,
       $W_{ij}^{(D)}$
       ,
           $W_{ij}^{(r,D)}$ 
           ,
            $W_{ij}^{(r_d)}$   
           and
        $W_{ij}^{(r_d,D)}$ increases the number of packets received by the gateways from  flows $4$ and $5$ whose source nodes are not connected  directly to the gateways. Furthermore, we observe that using parameters   $W_{ij}^{(r,D)}$  and
        $W_{ij}^{(r_d,D)}$ reduces the number of   received packets at  the gateways compared to using parameters $W_{ij}^{(r)}$,
       $W_{ij}^{(D)}$
      and
            $W_{ij}^{(r_d)}$.  

In Fig. \ref{fig4}, we observe that using   $W_{ij}^{(r)}$  improves the delay of all traffic flows which have high delay using 
 $W_{ij}$  parameter. But, using the parameters $W_{ij}^{(D)}$ and $W_{(i,j)}^{r_d}$  improves the delay of only some of those flows. Moreover, we observe that using  parameters $W_{ij}^{(r,D)}$ and $W_{ij}^{(r_d,D)}$ increases the average delay of packets for all the flows.

So far we investigated fairness using intuitive metrics. Now, we use the Jain's fairness index to indicate the amount of  fairness more accurately. If $x_1,\dots,x_n$ be the samples of  random variable $X$ and $x_i >0 , i=1, \dots,n$, then the Jain's fairness index is defined as \cite{jain}:
\begin{equation}
JFI=\frac{\left(\sum\limits_{i=\mathrm{1}}^{n}{x_i}\right)^\mathrm{2}}{n\sum\limits_{i=\mathrm{1}}^{n}{x_i^\mathrm{2}}},
\label{eq20}
\end{equation} 
where $0<JFI\leq 1$ and $JFI=1$ indicate the  complete fairness. Replacing $x_i$s in equation \eqref{eq20} with the desired parameters, the amount of the fairness corresponding to each of these parameters could be obtained. Here, we would like to investigate the fairness from throughput and delay points of view; then we substitute the parameters such as the throughput of  each flow, the average delay of each flow and the ratio of the throughput  and the average delay  of each flow  in \eqref{eq20}. Then, we obtain the amount of the fairness among traffic flows by running JGRP algorithm on the
network depicted in Fig \ref{fig1} based on parameters   $W_{ij}$
  ,
    $W_{ij}^{(r)}$
    ,
       $W_{ij}^{(D)}$
       ,
           $W_{ij}^{(r,D)}$ 
           ,
            $W_{ij}^{(r_d)}$   
           and
     $W_{ij}^{(r_d,D)}$ as shown in Table \ref{tab3}. From this table, we observe that  Jain's fairness index corresponding to throughput for   $W_{ij}^{(r,D)}$ and  $W_{ij}^{(r_d,D)}$ is equal to one, which means that  employing these parameters, complete fairness  will be provided among  flows. Comparing JFI corresponding to the average delay of flow packets, we observe that  all parameters $W_{ij}^{(r)}$, $W_{ij}^{(D)}$,  $W_{ij}^{(r,D)}$ , $W_{ij}^{(r_d)}$ and  $W_{ij}^{(r_d,D)}$ could improve the fairness from delay point of view, where   $W_{ij}^{(r,D)}$ and  $W_{ij}^{(r_d,D)}$ obtain the most amount of  fairness. By considering the ratio of the throughput and the average delay of each flow  as the parameter of Jain's fairness index, we can compare the fairness from both of the throughput  and delay point of views as shown in the third row of Table \ref{tab3}. We observe that  parameters $W_{ij}^{(r,D)}$ and  $W_{ij}^{(r_d,D)}$ could provide the most amount of fairness.
\begin{table*}[]
\begin{center}
\caption{the  amount of Jain's fairness index obtained among the flows with the implementation of  JGRP  algorithm on the network in Fig. \ref{fig1}}
\label{tab3}
\begin{tabular}{|c|c|c|c|c|c|c|}
\hline 
JGRP  algorithm  based on  & $W_{ij}$ &  $W_{ij}^{(r)}$ & $W_{ij}^{(D)}$ & $W_{ij}^{(r,D)}$ & $W_{ij}^{(r_d)}$ &  $W_{ij}^{(r_d,D)}$\\ 
\hline 

JFI  ($\text{per-flow throughput}$)&
$0.86$ & $0.99$ & $0.99$ & $1$& $0.99$ & $1$ \\ 
\hline 

JFI ($\frac{1}{\text{average delay of each flow }}$)
&$0.54$ & $0.72$ & $0.65$ & $0.78$& $0.71$ & $0.78$ \\
\hline 
JFI ($\frac{\text{per-flow throughput}}{\text{average delay of each flow}}$)
&$0.46$ & $0.67$ & $0.57$ & $0.75$& $0.66$ & $0.75$ \\
\hline
\end{tabular} 
\end{center}
\end{table*}
 %\begin{figure}[h]
 %\centering
%\includegraphics*[width=1\columnwidth]{allpacket.pdf}
%\caption{}
%\label{fig5}
%\end{figure}

Using parameters $W_{ij}^{(r,D)}$ and  $W_{ij}^{(r_d,D)}$ although improve the fairness in comparison with using   $W_{ij}$ parameter, increase the average delay and reduce the aggregate throughput and the number of packets received by the gateways. Therefore, in order to justify the performance of this algorithm, for each flow, , we show the ratio of the number of packets received by each gateway to the number of packets sent to that gateway for each flow in  Tables \ref{tab4} and \ref{tab5}. In Table \ref{tab4}, we observe that by using  $W_{ij}$, $W_{ij}^{(r)}$, $W_{ij}^{(D)}$ and
 $W_{ij}^{(r_d)}$, gateway $1$  almost does not receive  the packets of flows which their source nodes is connected directly to gateway 2, but  by using $W_{ij}^{(r,d)}$ and $W_{ij}^{(r_d,D)}$ these packets are  received more. This fact is also true for nodes with direct link  to gateway 1 as shown in Table \ref{tab5}. This is because the mentioned parameters  consider the rate of  links and the delay of  packets simultaneously, and this leads to the result that the intermediate links be activated and all the flows be communicated to all the gateways. Due to the multi-hop distance between the nodes and the non-adjacent gateway, the activation of intermediate links increases the average delay and decreases the aggregate throughput. 
\begin{table*}[]
 \begin{center}
 \caption{ Number of packet received  from each flow  by  the gateway 1 to the number of packets sent  by each flow to gateway 1 with the implementation of  JGRP  algorithm on the network Fig. \ref{fig1} }
  \label{tab4}
\begin{tabular}{|c|c|c|c|c|c|c|c|c|}
\hline
flow index(source node) & $1 (2)$& $2(3)$&$3(4)$ & $4 (5)$&$5 (6)$ & $6(7)$ &$7 (8)$& $8 (9)$
\\
\hline
 based on $W_{ij}$ &  $99.3$&	$96.02$&$	95.43$&	$99.63$&$18.58$&	$0$&$	
0$&$0$
\\
\hline
based on $W_{ij}^{(r)}$ & $98.45$&	$96.58$&$	96.29$&	$74.45$&	$86.59$&	$0$&	$0$&	$7.76$
\\
\hline
based on $W_{ij}^{(D)}$ & $98.32$&	 $96.52$ 
 & $97.12$
 	&	$82.49$
  & $89.09$&
 	$0$&
 	$0$&$54$
\\
\hline
based $W_{ij}^{(r,D)}$& $97.51$ &	$95.72 $&$	97.27 $&	 $76.53 $&$	89.18 $&	$19.49 $&$	19.75 $&$	51.26$
\\
\hline
based on $W_{ij}^{(r_d)}$& $98.39$ &	$96.47 $&$	96.06$&	 $	62.9$&	$86.37 $&	
 $0 $&	$0 $&
$35.17$
\\
\hline
based on $W_{ij}^{(r_d,D)}$ & $96.77 $ &	$ 94.85$&	$96.66$ &$	67.29$&	 $88.27$&$	18.79$&	$12.98$&	$54.12$
\\
\hline 
\end{tabular}
\end{center}
\end{table*}
\begin{table*}
 \begin{center}
 \caption{Number of packet received  from each flow  by  the gateway $2$ to the number of packets sent  by each flow to gateway $2$ with the implementation of  JGRP  algorithm on the network Fig. \ref{fig1} }
  \label{tab5}
\begin{tabular}{|c|c|c|c|c|c|c|c|c|}
\hline
flow index(source node) & $1 (2)$& $2(3)$&$3(4)$ & $4 (5)$&$5 (6)$ & $6(7)$ &$7 (8)4$& $8 (9)$
\\
\hline
 based on $W_{ij}$ &  $0$&$	0$&	$4.41$&$	90.6$&	$90.37$&	$99.22$	&$99.31$	&$95.47$
\\
\hline
based on $W_{ij}^{(r)}$ & $0.72	$&$0.35$&$	16.76$&	$92.49$&	$91.01	$&$98.45$&$	98.59$&	$96.51$
\\
\hline
based on $W_{ij}^{(D)}$ & $0$ &	$1.578$&$	16.47$&	$92.4$&$	86.30$&$	98.23$&	$98.8$&	$94.8$
\\
\hline
based on $W_{ij}^{(r,D)}$&$22.12$&$	39.01$&	$50.76$&	$90.58$&	$83.2$&	$97.09$&	$97.83$&	$95.11$
\\
\hline
based on $W_{ij}^{(r_d)}$& $2.78$&	$14.88$&	$38.93$&	$91.66$&	$84.94$&	$98.55$&	$98.38$&	$95.85$
\\
\hline
based on $W_{ij}^{(r_d,D)}$ & $30.66$&	$45.76$&$	52.85$&	$86.11$&	$73.65$&$	97.03$&	$97.38$&	$93.4$
\\
\hline 
\end{tabular}
\end{center}
\end{table*}

Now, we investigate the performance of the JGRP algorithm in networks with different sizes by running it based on all the parameters on the networks with $10$, $15$ and $20$ nodes over $2600$ time slots. For simulating networks with $15$ and $20$ nodes, we respectively distribute $15$ and $20$ nodes in a $450m\times450m$ and $500m\times500m$ square areas uniformly. Then, we form the topology of each of the networks using Algorithm \ref{algo1}. In order to obtain the simulation results of the networks with $15$ and $20$ nodes, we randomly select eight nodes as source nodes of the flows. We compare the simulation results of the networks with $10$, $15$ and $20$ nodes as shown in Tables \ref{tab6}, \ref{tab7}, \ref{tab8} and \ref{tab9}.

In Table \ref{tab6}, we observe that JGRP  algorithm based on $W_{ij}$ obtains the most aggregated throughput in the networks wit all three $10$, $15$ and $20$ nodes. Moreover, we observe that while increasing the number of nodes, the aggregated throughput   obtained by  JGRP  algorithm based on $W_{ij}$ is not reduced. But, in  JGRP  algorithm based on the other parameters, the aggregated throughput  is reduced when the number of the nodes is increased. This subject has two reasons; the first reason is that by using  parameters  $W_{ij}^{(r)}$, $W_{ij}^{(D)}$,  $W_{ij}^{(r,D)}$, $W_{ij}^{(r_d)}$ and $W_{ij}^{(r_d,D)}$, in each time  slot, the packet delay and the total allocated rate to the links are considered to fairly allocate the rates to the links. The second reason is that the number of nodes in the network is more than the number of the flows. These reasons lead to the fact that the links whose nodes are not the sources of the flows or a gateway, are likely to be activated. Then we have more packets remaining in the flow source nodes and consequently the generated traffic is reduced.
\begin{table}[]
\caption{Comparison of the  aggregated throughput for JGRP  algorithm based on  parameters  $W_{ij}$, $W_{ij}^{(r)}$, $W_{ij}^{(D)}$,  $W_{ij}^{(r,D)}$, $W_{ij}^{(r_d)}$ and $W_{ij}^{(r_d,D)}$ in the networks with $10$, $15$ and $20$ nodes.}
\label{tab6}
\centering
\begin{tabular}{|c|c|c|c|c|c|c|}
\hline 
 & $W_{ij}$& $W_{ij}^{(r)}$& $W_{ij}^{(D)}$ & $W_{ij}^{(r,D)} $& $W_{ij}^{(r_d)}$ & $W_{ij}^{(r_d,D)}$  \\ 
\hline 
$10$ nodes& $3.448$& $3.076 $& $3.213$&$ 2.915$ & $2.995$ &$2.785$\\ 
\hline 
$15$ nodes&  $3.707$& $3.024 $ & $3.268$ & $2.874$ &$2.941$ & $2.797$\\ 
\hline 
$20$ nodes & $3.414$&$2.887$  & $3.095$ & $2.781$ &$2.836$  &$2.837$ \\ 
\hline 
\end{tabular} 
\end{table}

In Table \ref{tab7}, we observe that by increasing the number of nodes, the average delay of one packet in the JGRP  algorithm based on all parameters increases. This subject is because of that when the  number of the nodes  is increased,  the number of hops between  the flow  source nodes and  the gateways could be increased.
\begin{table}[h]
\caption{Comparison of the average delay of one packet for the JGRP  algorithm based on  parameters  $W_{ij}$, $W_{ij}^{(r)}$, $W_{ij}^{(D)}$,  $W_{ij}^{(r,D)}$, $W_{ij}^{(r_d)}$ and $W_{ij}^{(r_d,D)}$ in the networks with $10$, $15$ and $20$ nodes.}
\label{tab7}
\centering
\begin{tabular}{|c|c|c|c|c|c|c|}
\hline 
 & $W_{ij}$& $W_{ij}^{(r)}$& $W_{ij}^{(D)}$ & $W_{ij}^{(r,D)} $& $W_{ij}^{(r_d)}$ & $W_{ij}^{(r_d,D)}$  \\ 
\hline 
$10$ nodes&$427.45$ &$368.43$ &$ 377.29$ &$437.94$ & $406.5$ & $476.05$ \\ 
\hline 
$15$ nodes& $400.94$ & $531.45$ & $ 468.13$&$656.37$&$576.7$  &$694.44$ \\ 
\hline 
$20$ nodes &  $552.32$ & $712.01$ &  $550.66$&$712.02$&$776 $ &$775.34$ \\ 
\hline 
\end{tabular} 
\end{table}

In Table \ref{tab8}, it can be observed  that increasing the number of the nodes in the networks, the number of  the packets received by the gateways is reduced for all the parameters, But, this reduction is more pronounced for $W_{ij}^{(r)}$, $W_{ij}^{(D)}$,  $W_{ij}^{(r,D)}$ , $W_{ij}^{(r_d)}$ and  $W_{ij}^{(r_d,D)}$. It can also be observed that increasing the number of nodes, the difference between the number of the received packets by the gateways using $W_{ij}$ and the other parameters increases. This  subject is because the number of  the flows is less than the number of nodes, then the attempt in the other parameters for obtaining the fairness between the links, leads to the result that the packets do not arrive at the  destination gateway.
\begin{table}[h]
\caption{Comparison of the total number of packets received by the gateways for the JGRP  algorithm based on  parameters  $W_{ij}$, $W_{ij}^{(r)}$, $W_{ij}^{(D)}$,  $W_{ij}^{(r,D)}$, $W_{ij}^{(r_d)}$ and $W_{ij}^{(r_d,D)}$ in the networks with $10$, $15$ and $20$ nodes.}
\label{tab8}
\centering
\begin{tabular}{|c|c|c|c|c|c|c|}
\hline 
 & $W_{ij}$& $W_{ij}^{(r)}$& $W_{ij}^{(D)}$ & $W_{ij}^{(r,D)} $& $W_{ij}^{(r_d)}$ & $W_{ij}^{(r_d,D)}$  \\ 
\hline 
$10$ nodes&$7248$ &  $6325$& $6618$ & $5685$& $6032$ &$5219$ \\ 
\hline 
$15$ nodes& $7557$ &$5440$& $6233$ &$4873$  &  $4983$&  $4455$  \\ 
\hline 
$20$ nodes &$7070$ &  $4107$&$5549$ &  $3340$&$3588 $ &$3588$ \\ 
\hline 
\end{tabular} 
\end{table}

In Table \ref{tab9}, we observe that increasing or decreasing the fairness does not have any direct  relation with the increase in the number of the nodes. The reason is that  the factors that affects the fairness are the allocated rate to the links and the number of hops between the source nodes and the  gateways.
\begin{table}[h]
\caption{Comparison of the  amount of Jain's fairness index ($\frac{\text{per-flow throughput}}{\text{average delay of each flow}}$) for the JGRP  algorithm based on parameters  $W_{ij}$, $W_{ij}^{(r)}$, $W_{ij}^{(D)}$,  $W_{ij}^{(r,D)}$, $W_{ij}^{(r_d)}$ and $W_{ij}^{(r_d,D)}$ in the networks with $10$, $15$ and $20$ nodes.}
\label{tab9}
\centering
\begin{tabular}{|c|c|c|c|c|c|c|}
\hline 
 & $W_{ij}$& $W_{ij}^{(r)}$& $W_{ij}^{(D)}$ & $W_{ij}^{(r,D)} $& $W_{ij}^{(r_d)}$ & $W_{ij}^{(r_d,D)}$  \\ 
\hline 
$10$ nodes&$0.48$ &$ 0.71 $&$ 0.57$ & $0.73$ & $0.72 $& $0.75$ \\ 
\hline 
$15$ nodes& $0.47$ &$0.62$  &$0.6$  & $0.66$ &$0.64$  & $0.68$ \\ 
\hline 
$20$ nodes &$0.65$ &$0.87$  & $0.86$ &$ 0.93$ & $0.89$ & $0.9$\\ 
\hline 
\end{tabular} 
\end{table}
\subsection{Multi-Radio Multi-Channel  JGRP algorithm (MR-MC JGRP)}
Now, we introduce MR-MC JGRP as an extension of the JGRP algorithm for multi-radio multi-channel networks. In the special case where the number of radios on each node is equal to the number of frequency channels, the extension is very simple. For this purpose, in the \ref{algo20-3}$^{\rm{th}}$ step of Algorithm \ref{algo3}, partition the set of schedules to some subsets and in each of them, use one of the frequency channels. Then solve problem \eqref{eq9} for each of the frequency channels individually.

\begin{figure}[t]
 \centering
\includegraphics*[width=.5\columnwidth]{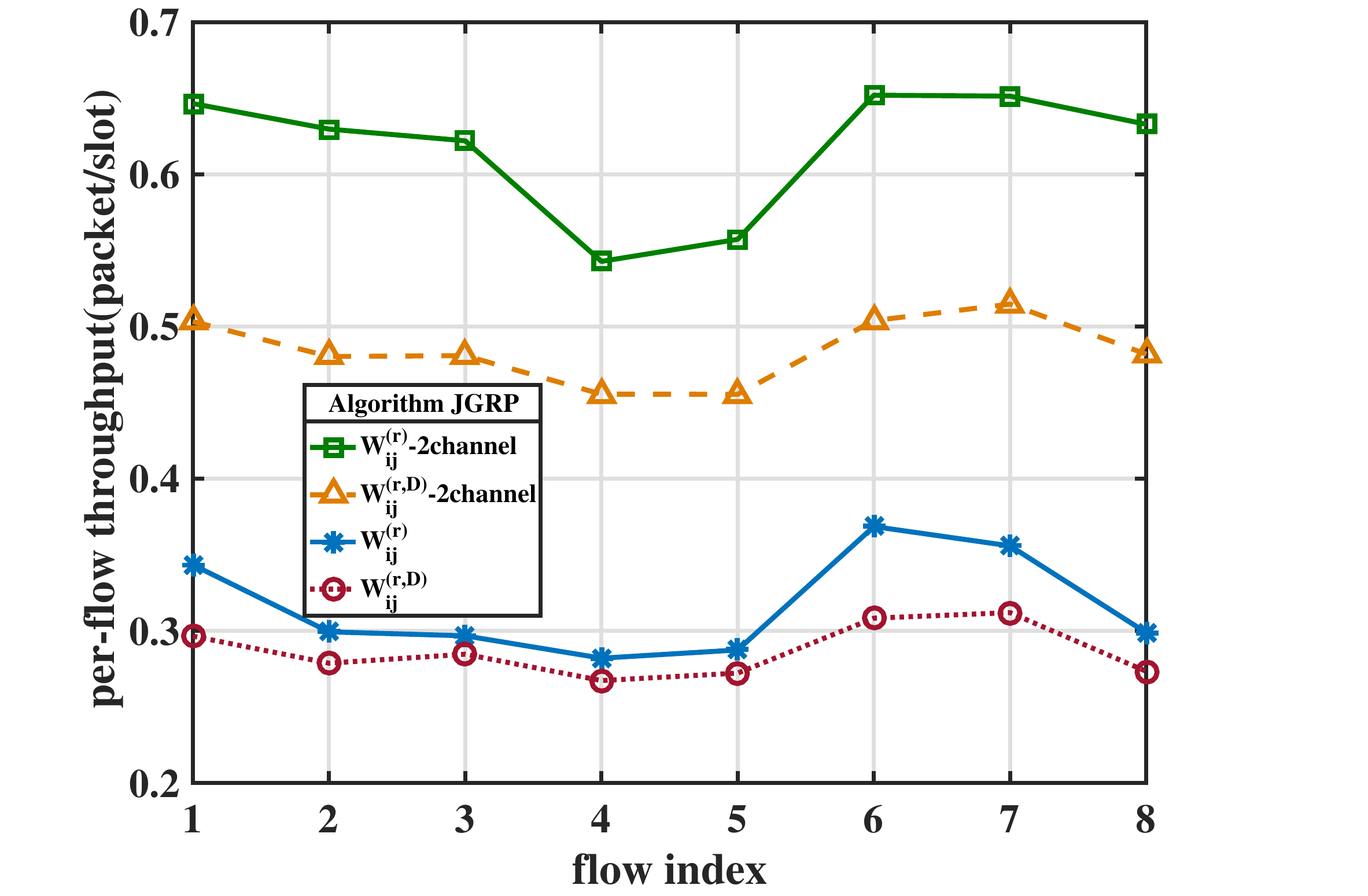}
\caption{Throughput of each flow in MR-MC JGRP algorithm based on parameters  $W_{ij}^{(r)}$ and  $W_{ij}^{(r,D)}$.}
\label{fig6}
\end{figure}

\begin{figure}[t]
 \centering
\includegraphics*[width=.5\columnwidth]{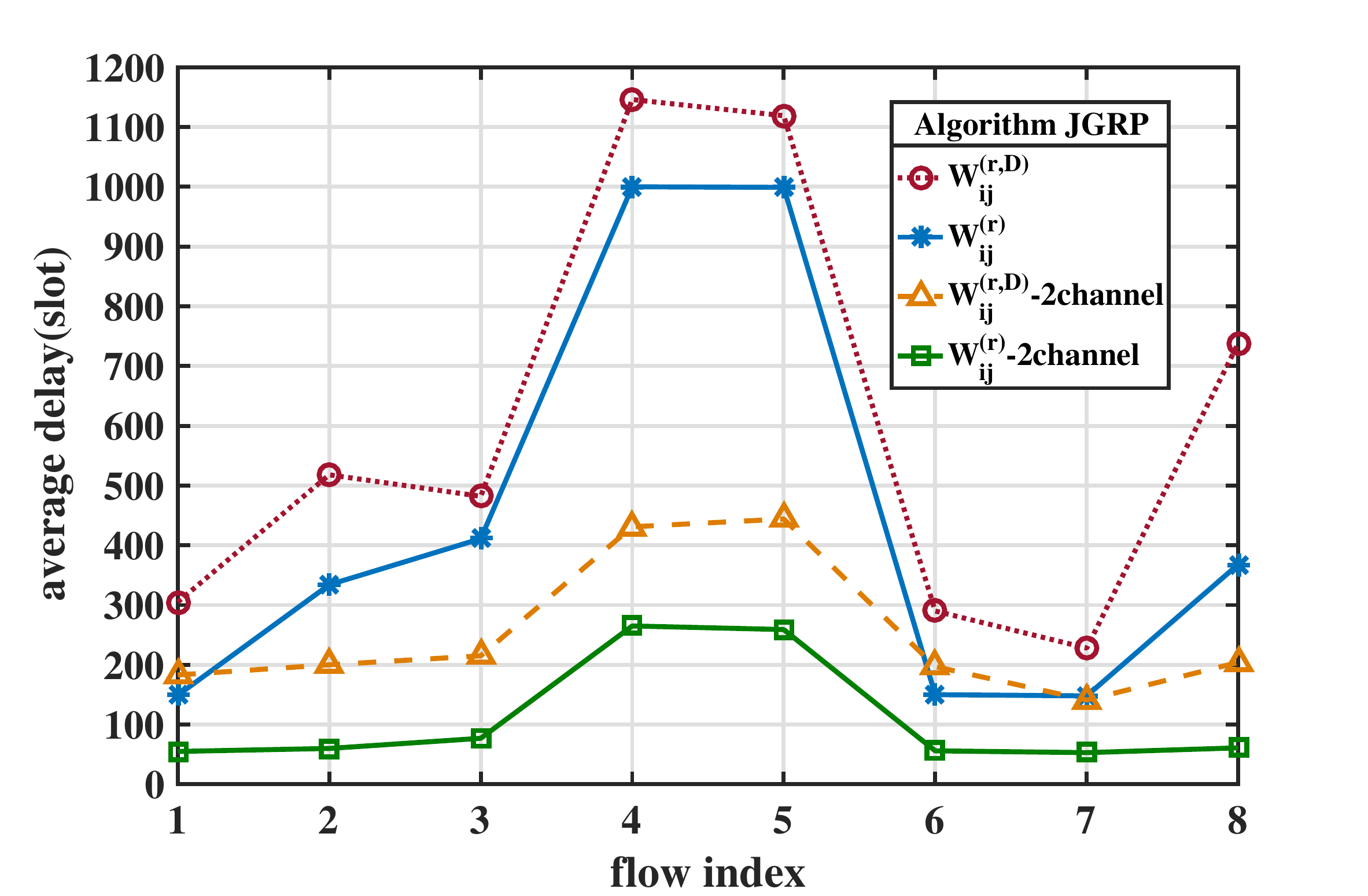}
\caption{Average delay of one packet for each flow in MR-MC JGRP algorithm based on parameters   $W_{ij}^{(r)}$ and  $W_{ij}^{(r,D)}$.}
\label{fig7}
\end{figure}

For simulation, we assume that each network node has two radios and the number of accessible frequency channels equals to the number of radios. Then, we run the JGRP  algorithm based on $W_{ij}^{(r)}$ and $W_{ij}^{(r,D)}$  over $10^4$ time slots and compare this  with the one radio one frequency channel case. The results are shown in Figs \ref{fig6} and \ref{fig7}.

In Fig. \ref{fig6}, we observe that in the JGRP algorithm based on  both parameters $W_{ij}^{(r)}$ and $W_{ij}^{(r,D)}$  using two frequency channels doubles the throughput of the flows comparing with the case of one radio one frequency channel.

In Fig. \ref{fig7}, we observe that in the JGRP algorithm based on  both parameters  $W_{ij}^{(r)}$ and $W_{ij}^{(r,D)}$ using  two frequency channels and  two radios per node reduces the average delay and improves the fairness from the delay point of view.

\section{Conclusion}
\label{sect7}
In this paper, we formulated the problem of network utility maximization for multiple gateways wireless mesh networks by considering joint rate control, traffic splitting, scheduling, routing, link rate allocation and power control assuming SINR interference model. In addition, by considering the complexity of this problem, we proposed the JGRP algorithm as a sub-optimal solution. Then, in order to improve the fairness, we defined $W_{ij}^{(r)}$, $W_{ij}^{(D)}$,  $W_{ij}^{(r,D)}$, $W_{ij}^{(r_d)}$ and $W_{ij}^{(r_d,D)}$ as new parameters to be used in the JGRP algorithm instead of the differential backlog ($W_{ij}$). Simulation results illustrate that using parameters $W_{ij}^{(r)}$ and $W_{ij}^{(r_d)}$ in which the sum of the allocated link rates are considered in their definition, improved the fairness from  average delay and throughput points of view. Moreover, by using these parameters, the number of packets received by the gateways is not reduced and the average delay is not increased in comparison to
 $W_{ij}$ parameter . We also showed that using $W_{ij}^{(r,D)}$ and $W_{ij}^{(r_d,D)}$ in which the sum of the allocated link rates  and the delay of packets were considered in their definition, not only  improves the fairness, but also improves the communication between mesh nodes and gateways. However, its negative effects increase the average delay and decrease the number of packets received by the gateways in comparison with other parameters. Finally, we extended the JGRP algorithm for multi-radio multi-channel networks in which the number of radios per node is equal to the number of accessible frequency channels.  We showed that using two frequency channels and radios when   $W_{ij}^{(r)}$ and $W_{ij}^{(r,D)}$ are used doubles the throughput of the flows and reduces the average delay in comparison to the single radio single channel case. 

\bibliographystyle{ieeetr}
\bibliography{mybibfile}
% biography section
% 
% If you have an EPS/PDF photo (graphicx package needed) extra braces are
% needed around the contents of the optional argument to biography to prevent
% the LaTeX parser from getting confused when it sees the complicated
% \includegraphics command within an optional argument. (You could create
% your own custom macro containing the \includegraphics command to make things
% simpler here.)
%\begin{IEEEbiography}[{\includegraphics[width=1in,height=1.25in,clip,keepaspectratio]{mshell}}]{Michael Shell}
% or if you just want to reserve a space for a photo:

% You can push biographies down or up by placing
% a \vfill before or after them. The appropriate
% use of \vfill depends on what kind of text is
% on the last page and whether or not the columns
% are being equalized.

%\vfill

% Can be used to pull up biographies so that the bottom of the last one
% is flush with the other column.
%\enlargethispage{-5in}

% that's all folks
\end{document}